\documentclass[journal]{IEEEtran}

\usepackage[utf8]{inputenc} 
\usepackage[T1]{fontenc}
\usepackage{color}
\usepackage{float}
\usepackage{amsbsy}
\usepackage{amstext}
\usepackage{setspace}
\usepackage{orcidlink}
\usepackage{graphicx}
\usepackage{bm}
\usepackage{float}
\usepackage{amsmath, amssymb}
\usepackage{siunitx}
\usepackage{gensymb}
\usepackage{upgreek}
\usepackage{breqn}
\usepackage{blindtext}
\usepackage{subcaption}
\usepackage{widetext}

\begin{document}

\title{
 Second-Harmonic Generation at a Fourth-Order Exceptional Point Degeneracy
}

\author{Albert~Herrero-Parareda~\orcidlink{https://orcid.org/0000-0002-8501-5775},
        Domenico de Ceglia~\orcidlink{https://orcid.org/0000-0001-5736-6298}, Maria Antonietta Vincenti~\orcidlink{https://orcid.org/0000-0001-6698-2973}, Attilio Zilli~\orcidlink{https://orcid.org/0000-0003-1845-6850}, Maxim R. Shcherbakov~\orcidlink{https://orcid.org/0000-0001-7198-5482},
        and~Filippo~Capolino~\orcidlink{https://orcid.org/0000-0003-0758-6182}
\thanks{Albert Herrero-Parareda, Attilio Zilli, Maxim R.\,Shcherbakov and Filippo Capolino are affiliated with the Department of Electrical Engineering and Computer Science, University of California, Irvine, CA, 92697 USA e-mail: f.capolino@uci.edu.}
\thanks{Domenico de Ceglia and Maria Antonietta Vincenti are affiliated with the Dipartimento di Ingegneria dell'Informazione, Università degli Studi di Brescia, Italy.}
\thanks{Attilio Zilli is affiliated with the Dipartimento di Fisica, Politecnico di Milano, Italy.}
}

\maketitle

\begin{abstract}

An anomalous flat-band dispersion provided by a degenerate band edge (DBE) of longitudinal optical modes in a double-grating waveguide is used to enhance second-harmonic generation (SHG). The DBE is a fourth-order exceptional point degeneracy (EPD) in a lossless and gainless waveguide, characterized by the coalescence of four eigenmodes that establish a frozen mode in a cavity. At a DBE resonance, the cavity quality factor scales $Q\propto N^5$, where $N$ is the number of unit cells of the grating waveguide. In our numerical experiments, we observe the peak intensity of the fundamental field in the edge-excited cavity scaling as $I_1\propto N^{3.6}$. This leads to a highly efficient SHG process that is radiated vertically from the cavity (i.e., normal to the grating) without requiring collinear phase matching, with a conversion efficiency scaling as $\eta\propto N^{8.27}$.  These results establish DBE-based waveguides as promising platforms for miniaturized efficient nonlinear photonic devices.
\end{abstract}

\section{Introduction}
\label{ch:Intro}

The development of efficient nonlinear photonic integrated circuits is essential to advance technologies such as quantum information processing, optical signal processing, and ultrafast lasers~\cite{mobini_algaas_2022}. Second-harmonic generation (SHG), whereby two photons at a fundamental frequency $f_1$ convert into one at $f_2 = 2f_1$, is among the most relevant nonlinear optical processes \cite{boyd_nonlinear_2008}. However, SHG efficiency remains low in miniaturized devices due to weak nonlinear coefficients and poor modal overlap. Addressing this challenge requires new strategies to enhance field amplitudes in compact structures. To this end, researchers have explored photonic crystals and resonant cavities that localize optical fields \cite{shi_defective_2001, yariv_photonics_2007}. 

A key figure of merit in nonlinear photonics is the dependence of the conversion efficiency on the device length, because high-order scaling can enable strong conversion in miniaturized platforms. In Ref.~\cite{deAngelis_Maximum_2007}, an $N^8$ scaling with the number of unit cells $N$ is demonstrated in an ideal photonic crystal made of a periodic stack of layers under non-phase-matching conditions. However, it relies on the assumption that both $f_1$ and $f_2$ lie exactly at the first resonances of their respective band edges, even though resonances approach each band edge differently depending on the dispersion's curvature. Furthermore, in Ref.~\cite{deAngelis_Maximum_2007} the ideal photonic crystal was such that the upper band edge is exactly at twice the frequency of the lower band edge. These assumptions are difficult to realize in realistic designs.  

\begin{figure}[h]
    \centering
    \begin{subfigure}[t]{0.9\linewidth}
        \centering
        \includegraphics[width=\linewidth]{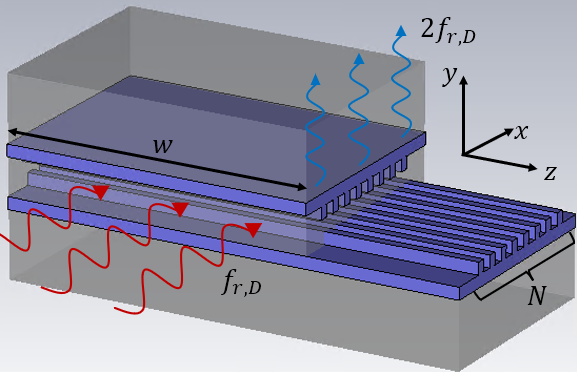}
        \caption{}
        \label{fig:SchematicDBESHG_a}
    \end{subfigure}
    \hfill
    \begin{subfigure}[t]{0.8\linewidth}
        \centering
        \includegraphics[width=\linewidth]
        {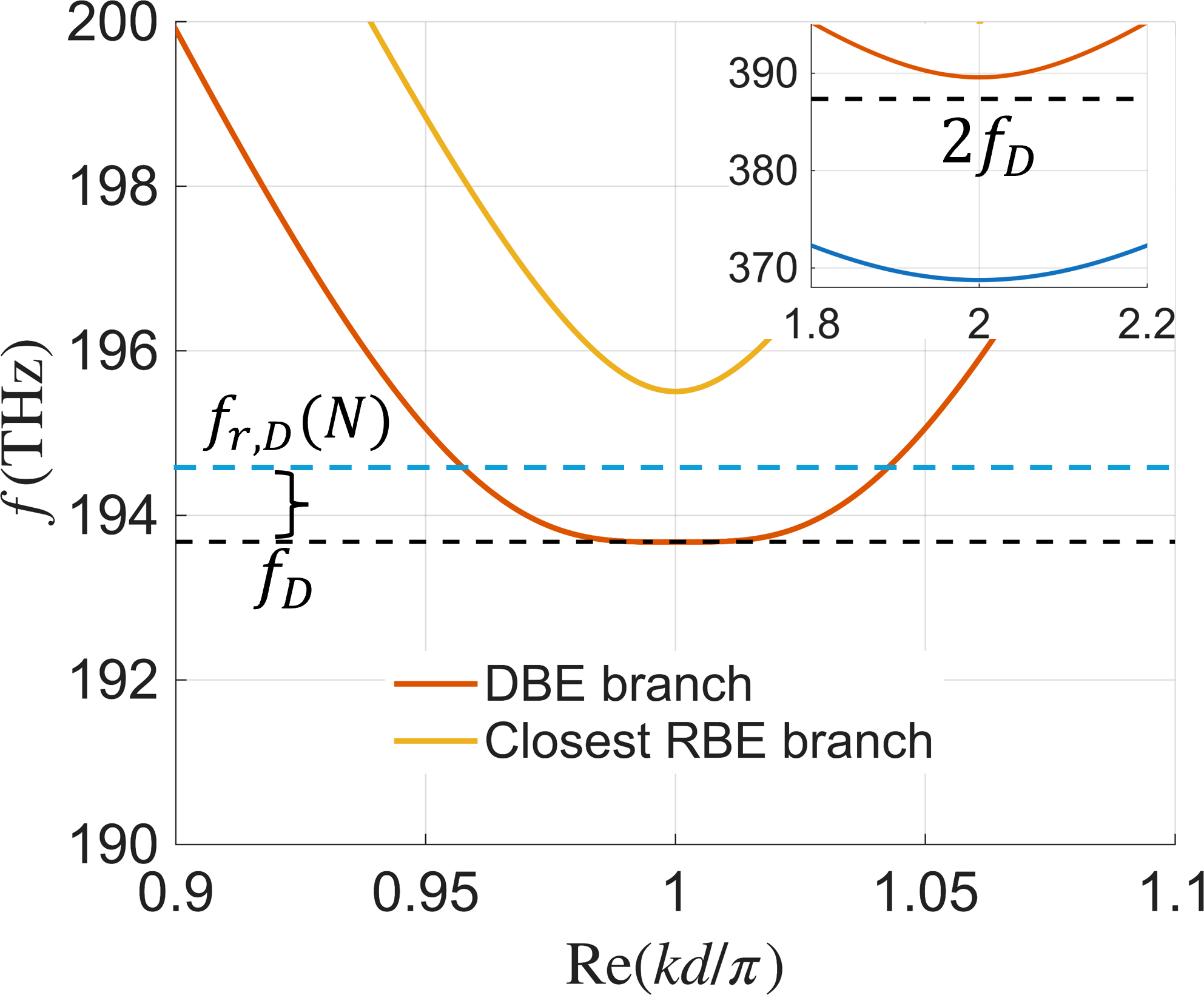}
        
        \caption{}
        \label{fig:Dispersions}
    \end{subfigure}
    \caption{(a) Schematic of the three-dimensional double-grating waveguide supporting a DBE 
    composed of $N$ unit cells along the $x$ direction and width $w$ in the transverse $z$ direction. The upper section is partially removed for visual clarity. A pump wave at the DBE resonance frequency $f_{r,D}$ propagates in the positive $x$ direction and generates a second harmonic at $2f_{r,D}$, which radiates in the $y$ direction, i.e., vertically. Collinear phase matching at $2f_{r,D}$ is not required, whereas the phase condition for vertically-emitted second-harmonic is satisfied when the pump is tuned at the DBE. (b) Dispersion diagram of $z$-polarized modes  near  the fundamental frequency, with the flat band at the Brillouin zone edge indicating the DBE at $f_D$ (black dashed). The bracket indicates the difference between the resonant frequency $f_{r,D}$ (blue dashed-line), and the DBE frequency $f_D$ (black-dashed line). The inset shows the dispersion diagram around $2f_D$ (black-dashed line). The SHG at $2f_{r,D}$ would not excite a mode near $\operatorname{Re}(kd/\pi)=2$.} 
    \label{fig:SchematicDBESHG}
\end{figure}

\begin{figure}
    \centering
    \includegraphics[width=0.8\linewidth]{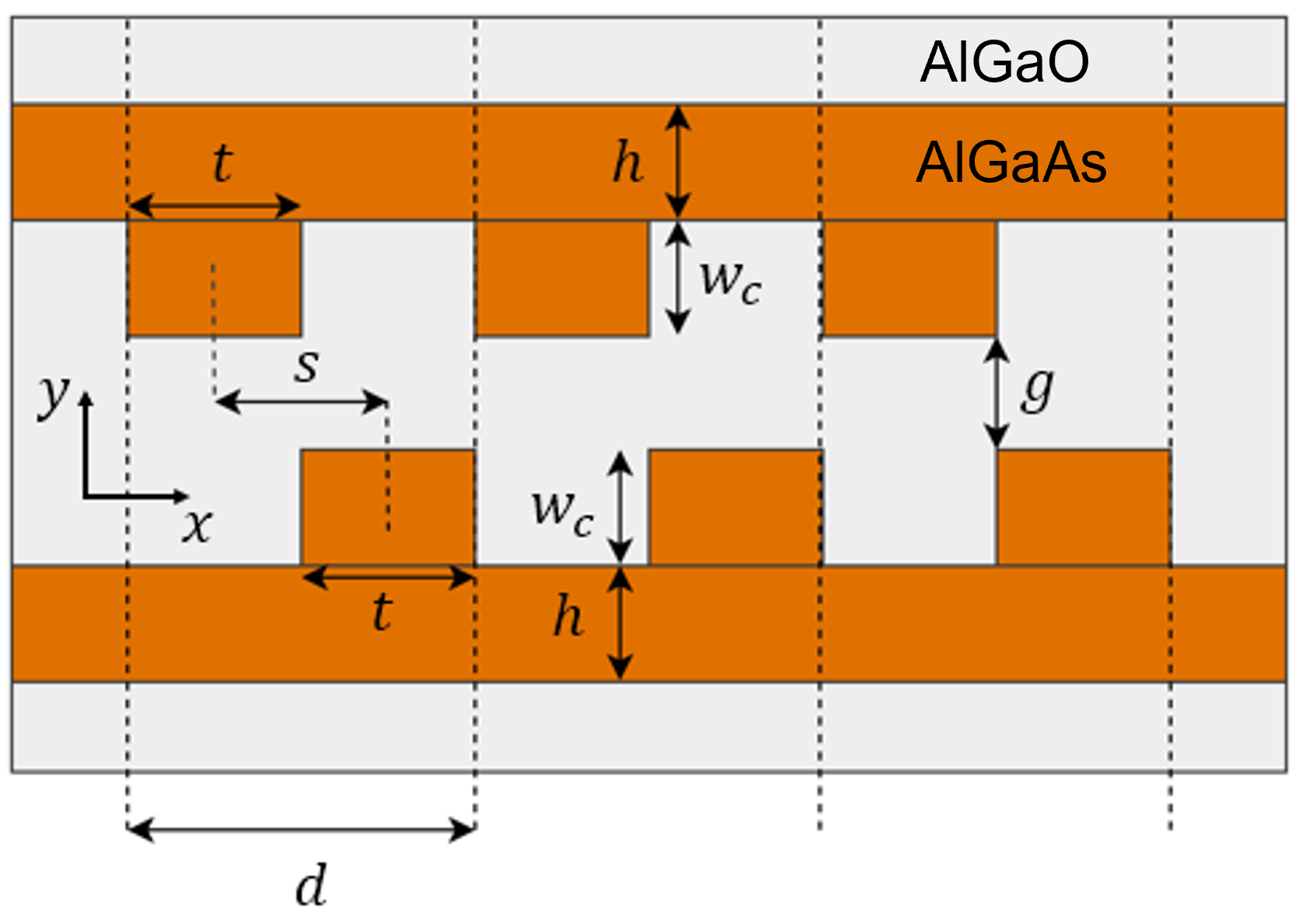}
    \caption{Transverse section of a double grating supporting a DBE of guided longitudinal modes in the $x$ direction, polarized along $z$ (out-of-plane). The two oppositely-facing gratings are shifted by a distance $s$ along $x$ to break the mirror symmetry, enabling the occurrence of the DBE that leads to a very flat band in the dispersion diagram provided in the first equation of \eqref{eq:DBERBEDisp}. Orange and gray areas indicate high-index (AlGaAs) and low-index (AlGaO) regions of the structure, respectively.}
    \label{fig:double_grating}
\end{figure}

Recently, bound states in the continuum (BICs) have drawn much interest for their ability to trap light without radiation losses, leading to strong field confinement and enhanced nonlinear interactions \cite{carletti_giant_2018,minkov_doubly_2019, ye_second_2022,yuqiu_highly_2023}. However, finite-size effects severely deteriorate BICs mode in practical applications \cite{hao_performance_2025}. Alternative mechanisms that exploit dispersion-engineered properties would offer new opportunities to enhance nonlinear effects.

In this work, we exploit the concept of exceptional point degeneracy (EPD) to enhance the nonlinear SHG efficiency. EPDs are unique points in a system where multiple eigenmodes coalesce (see for example Ref.~\cite{heiss_exceptional_2004, heiss_physics_2012}, or~\cite{figotin_slow_2006, figotin_slow_2011} where Figotin and Vitebskiy refer to EPDs as stationary points). EPDs in photonic systems can exist in gain and loss balanced systems, such as parity-time symmetric structures \cite{ruter_observation_2010, ramezani_unidirectional_2010, othman_theory_2017}, or in purely passive systems through dispersion engineering \cite{figotin_slow_2006, nada_theory_2017, furman_frozen_2023}. The structured resonance mechanism discussed here is based on an EPD belonging to this latter class, i.e., a fourth-order EPD in a periodic waveguide without loss and gain \cite{nada_theory_2017}, known as the degenerate band edge (DBE). At a regular band edge (RBE), a second-order EPD, two counter-propagating modes merge, leading to a standing wave with zero group velocity \cite{figotin_gigantic_2005}. A DBE occurs when two counter-propagating and two evanescent modes merge, forming a frozen mode. A DBE at optical frequencies has been experimentally demonstrated in silicon photonic waveguides through dispersion reconstruction and by observing the scaling of the quality factor $Q_D \propto N^5$ in finite-length structures \cite{burr_degenerate_2013, wood_degenerate_2015}. Further experimental validations exist at microwave frequencies \cite{othman_experimental_2017, abdelshafy_exceptional_2019,Mealy2020GeneralConditions, Zheng22MTTSynthMeasDBE}.

When operated near the DBE frequency, a finite-length waveguide exhibits strongly enhanced field amplitudes away from the cavity edges,  and a group delay that scales with the fifth power of the waveguide length \cite{figotin_slow_2011, wood_degenerate_2015, nada_theory_2017, othman_giant_2016, othman_experimental_2017, veysi_degenerate_2018}. In contrast, the group delay near a regular band edge (RBE) grows only with the cube of the waveguide length. The peak field intensities, defined as $I = |\mathbf{E}|^2$, scale asymptotically with the number of unit cells $N$ as

\begin{equation}
\begin{array}{c}
    \max(I_{\text{RBE}}) \propto N^2, \vspace{1em}\\
    \max(I_{\text{DBE}}) \propto N^4,
\end{array}
\label{eq:MaxFieldEnhancement}
\end{equation}

which we aim to exploit to achieve a DBE-based nonlinear conversion efficiency with a larger than $N^8$ asymptotic scaling, even without exciting Bloch modes at the second-harmonic frequency. To illustrate this effect, we design a periodic waveguide composed of two coupled gratings that support a DBE at the fundamental frequency. Following the approach of \cite{monat_investigation_2010} and \cite{bile2025}, the second-harmonic signal is allowed to radiate vertically, i.e., out of the plane of the structure, namely, perpendicular to the propagation direction of the guided mode at the fundamental wavelength. This open configuration facilitates practical implementations. Figure~\ref{fig:SchematicDBESHG_a} shows the three-dimensional concept: a waveguide made of $N$ periodic unit cells along the $x$ direction, with a finite width $w$ in the transverse $z$ direction. A pump at the DBE resonance frequency $f_{r,D}$ (the closest resonance to the DBE frequency) travels along the positive $x$ direction. The resulting frozen-mode intensity enhancement leads to efficient SHG, with the generated signal at $2f_{r,D}$ radiating out of the waveguide in the $y$ direction. By leveraging the frozen mode effect, we achieve high field intensities at the fundamental frequency ($I_1$) deep inside the cavity that are shown to scale as $N^{3.6}$, yielding a nonlinear conversion efficiency that scales as $N^{8.27}$.

Figure~\ref{fig:Dispersions} shows the real-wavenumber branches of the modal dispersion diagram of the double grating, where the DBE appears as a flat region (orange) near the Brillouin zone edge at $\operatorname{Re}(kd/\pi)=1$. The black dashed line represents the DBE frequency $f_D$, whereas the bracket highlights the narrow spectral distance at which the frozen mode resonance $f_{r,D}$ in a finite-length grating occurs just above the DBE frequency. The inset displays the real part of the modal wavenumbers near the edge of the second Brillouin zone, confirming the absence of Bloch modes at twice the DBE frequency $f_D$ (shown as a dashed black line) near $\operatorname{Re}(kd/\pi)=2$.

Our results highlight the potential of combining EPDs with nonlinear processes to overcome current limitations in integrated photonic devices. The proposed strategy provides a scalable and practical solution for enhancing nonlinear interactions, enabling an anomalous scaling of the frequency conversion efficiency in photonic integrated circuits.

The paper is organized as follows. The AlGaAs-based double-grating design that supports a DBE at the fundamental frequency is introduced in Section~\ref{ch:Design}, and we analyze its performance in Section~\ref{ch:FF}. Section~\ref{ch:shg} presents the features of the DBE-enhanced SHG process. The results are summarized in Section~\ref{ch:Conclusions}.

\section{Design of the DBE-supporting double grating}
\label{ch:Design}

The model structure chosen to enhance the field and the nonlinear conversion efficiency consists of two facing periodic gratings, as shown in Figure~\ref{fig:double_grating} and initially presented in Ref.~\cite{mealy_mirror_2023}. The structure is created by applying a mirror operation with respect to the $(x,z)$ plane, followed by a longitudinal translation $s$ along $x$ of one grating. The coupling forms a system that supports four guided modes. To obtain a DBE in this structure, its four modes must coalesce with the same frequency, Bloch wavenumber, and polarization. The translation $s$ breaks the mirror symmetry of the device, a necessary condition to obtain coalescence between the four modes supported by this structure \cite{mealy_mirror_2023}. This waveguide can be designed using several material platforms (silicon-on-insulator technology, $\text{SiN}$, $\text{InP}$, $\text{LiNbO}_3$, $\text{GaAs}$, etc.). Here, we choose a core made of Al$_{0.2}$Ga$_{0.8}$As, whose bandgap corresponds to a wavelength of approximately 750\,nm (or 400\,THz) \cite{adachi_gaas_1994}. Since our target SHG wavelength ($\approx$\,775\,nm) lies at longer wavelength, it falls in the transparency region of the material, where interband absorption losses are negligible. In addition, Al$_{0.2}$Ga$_{0.8}$As provides a large nonlinear coefficient, $d_{14} = 125$\,pm/V. Around the fundamental frequency $f_1\approx 193.68$\,THz, we have a core refractive index of $n_{\mathrm{c},1} = 3.2837$, and $n_{\mathrm{c},2} = 3.5607$ at the SHG frequency $f_2 \approx 387.36$\,THz \cite{papatryfonos_refr_2021}. The cladding is AlGaO with a refractive index of $n_{\mathrm{clad}} = 1.60$ \cite{choquette_advances_1997} at all studied frequencies. We account for the refractive index difference between $f_1$ and $f_2$ in the core, but neglect dispersion within the narrow frequency ranges of interest around each frequency.

We aim to design the double grating so that these four modes coalesce into the DBE-associated frozen mode. This occurs at the DBE frequency $f_D\approx 193.679$\,THz and at the edge of the first Brillouin Zone, i.e., $k_Dd=\pi$ (where $d$ is the unit cell period), leading to a high SHG nonlinear conversion efficiency $\eta$. This occurs because the DBE displays enhanced intensities in the cavity compared to the RBE as shown in Eq.~\eqref{eq:MaxFieldEnhancement}. Near these exceptional points, their dispersions are approximated as \cite{burr_degenerate_2013, othman_giant_2016, othman_experimental_2017, nada_theory_2017, nada_giant_2018, veysi_degenerate_2018}

\begin{equation}
    \begin{array}{c}
        (f - f_D) \approx  \frac{h_D}{2 \pi}(k - k_D)^4,
        \vspace{1em}\\
        (f - f_R) \approx  \frac{h_R}{2 \pi}(k - k_R)^2,
    \end{array}
    \label{eq:DBERBEDisp}
\end{equation}

where $f_{D}$, $f_{R}$ and $k_{D}$, $k_{R}$ are the EPD frequencies and the Bloch wavenumbers for the DBE and RBE cases, respectively. The parameters $h_D$ and $h_R$ describe the local flatness of the band edges and are sometimes referred to as "flatness parameters". The DBE displays a much flatter dispersion than the RBE in Figure~\ref{fig:Dispersions}, and its flatness is caused by the coalescence  of the four-way degenerate mode \cite{figotin_electromagnetic_2013}. At the DBE, one has the derivatives ${\mathrm d}^n\omega/{\mathrm d}k^n=0$ for $n=1,2,3$, and ${\mathrm d}^4\omega/{\mathrm d}_k^4= 24 h_D$. For the case presented in this paper, where the band gap is below $f_D$, one has $h_D>0$; otherwise, $h_D<0$. The difference between the DBE resonance frequency $f_{r,D}$ (the closest resonance to the DBE) and the DBE frequency $f_D$ determines the magnitude of the enhanced field amplitudes in the cavity, the linewidth of the resonance, its group velocity, and quality factor $Q_D$. 

The double-grating design that supports a DBE at frequency $f_D = 193.679$\,THz is obtained via a combination of manual and numerical parameter optimization. The key design metric is the minimization of the second derivative of the Bloch modes at the edge of the Brillouin zone at $k=\pi/d$, which is the first derivative of the group velocity, $v_{\text{g}} = {\mathrm d}\omega/{\mathrm d} k$. Specifically, the DBE condition is achieved when $|{\mathrm d}^2\omega/ {\mathrm d} k^2|_{k=\pi/d} \to 0$ which ensures that all four Bloch modes supported by the double grating become degenerate. In contrast, minimizing only the group velocity magnitude $|v_{\text{g}}|\to 0$ at $k=\pi/d$ may yield an RBE instead. The value of the geometric parameters of the double grating was also optimized so that the frequency difference between the DBE and the nearest RBE is large enough to prevent issues with the resonances of the finite-length structure. 

The double grating shown in Figure~\ref{fig:double_grating} is modeled in COMSOL Multiphysics, a full-wave electromagnetic solver based on finite-element method as detailed in Appendix B. Our objective is to obtain longitudinal modes that propagate in the $x$ direction and are polarized in the $z$ direction (using the coordinate system of Figures~\ref{fig:SchematicDBESHG_a} and \ref{fig:double_grating}). 

The modal dispersion diagram for the infinitely long double grating is computed by simulating a single unit cell of the waveguide using the eigenfrequency study in COMSOL Multiphysics, with periodic boundary conditions. Figure~\ref{fig:Dispersions} shows the resulting purely-real branch of the dispersion diagram. The optimization algorithm uses gradient descent methods to minimize the magnitude of the group velocity dispersion at $k = \pi/d$ and achieve the formation of the DBE by optimizing six geometric parameters indicated in Figure~\ref{fig:double_grating}: period $d$, height $h$, gap $g$, core width $w_c$, thickness $t$, and shift $s$. The resulting optimized waveguide parameters are (in nm): $d=300$, $h=150$, $g=120$, $w_c = 184$, $t=200$, and $s=100$. The solid-orange curve in Figure~\ref{fig:Dispersions} denotes the modes that coalesce at the DBE frequency $f_D = 193.679$\,THz, while the solid-yellow curve at higher frequencies denotes the modes that coalesce forming a regular band edge~(RBE) at $195.5$\,THz that also occurs in the same structure. 

\section{Linear response}
\label{ch:FF}

Next, we consider a finite double-grating cavity made of $N$ unit cells in the $x$ direction terminated by regular uniform waveguides with height $h$ (without adding any mirror) as illustrated on the sides of Figures~\ref{fig:SchematicDBESHG_a} and \ref{fig:double_grating}.The properties of the DBE-associated frozen mode are accessed through the closest Fabry--Pérot resonance to the DBE frequency, termed the "DBE resonance" $f_{r,D}$, as in \cite{burr_degenerate_2013, nada_theory_2017, nada_giant_2018, veysi_degenerate_2018}. Similarly, the closest resonance to an RBE is referred to as the "RBE resonance" $f_{r,R}$. These resonances converge to the corresponding EPD frequencies with increasing $N$, following the asymptotic expressions

\begin{equation}
\begin{array}{c}
    f_{r,D} \sim f_{D} + \frac{h_D}{2 \pi} \left(\frac{\pi-\varphi}{N d}\right)^4,   \vspace{1em}\\
    f_{r,R} \sim f_{R} + \frac{h_R}{2 \pi} \left(\frac{\pi}{N d}\right)^2,
    \end{array}
\label{eq:DBEresonFreqAsymp}
\end{equation}

The scaling of the DBE resonance approaches the DBE frequency with the $1/N^4$ asymptotic difference as shown in \cite{burr_degenerate_2013, othman_giant_2016, othman_experimental_2017, nada_theory_2017, nada_giant_2018, veysi_degenerate_2018, abdelshafy_exceptional_2019}, while the RBE resonances approach the RBE frequency with the $1/N^2$ asymptotic difference \cite{bendickson_analytic_1996}. The angle $\varphi$ accounts for the presence of the evanescent waves in a DBE resonant cavity \cite{figotin_gigantic_2005,  burr_degenerate_2013}.  (Note that also the term $\varphi$ should have been included in \cite{othman_experimental_2017, nada_theory_2017, veysi_degenerate_2018, abdelshafy_exceptional_2019}.)

\begin{figure*}[h]
    \centering
    \begin{subfigure}[t]{0.45\linewidth}
        \centering
        \includegraphics[width=\linewidth]{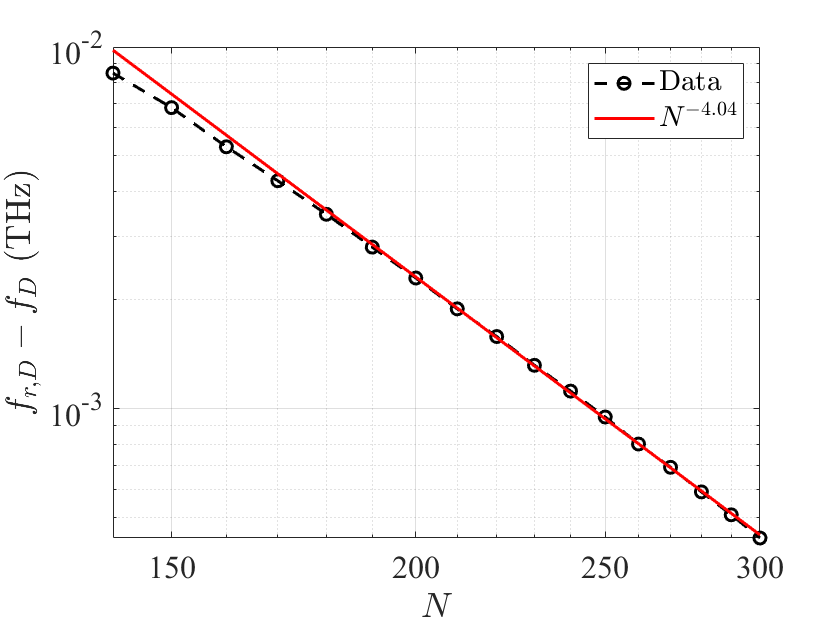}
        \caption{}
        \label{fig:ResvsN}
    \end{subfigure}
    \begin{subfigure}[t]{0.45\linewidth}
        \centering
        \includegraphics[width=\linewidth]{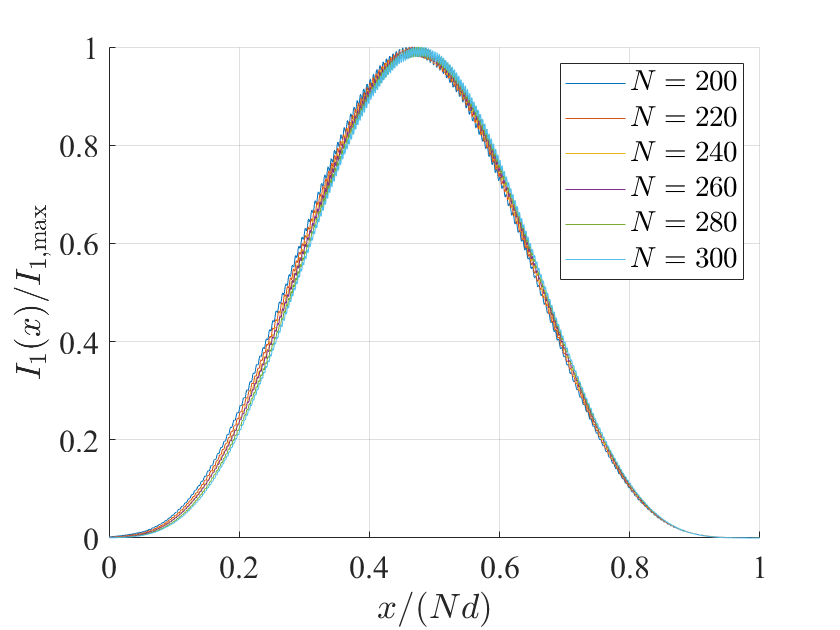}
        \caption{}
        \label{fig:normIntensityProfile}
    \end{subfigure}
    \hfill
    \begin{subfigure}[t]{0.45\linewidth}
        \centering
        \includegraphics[width=\linewidth]{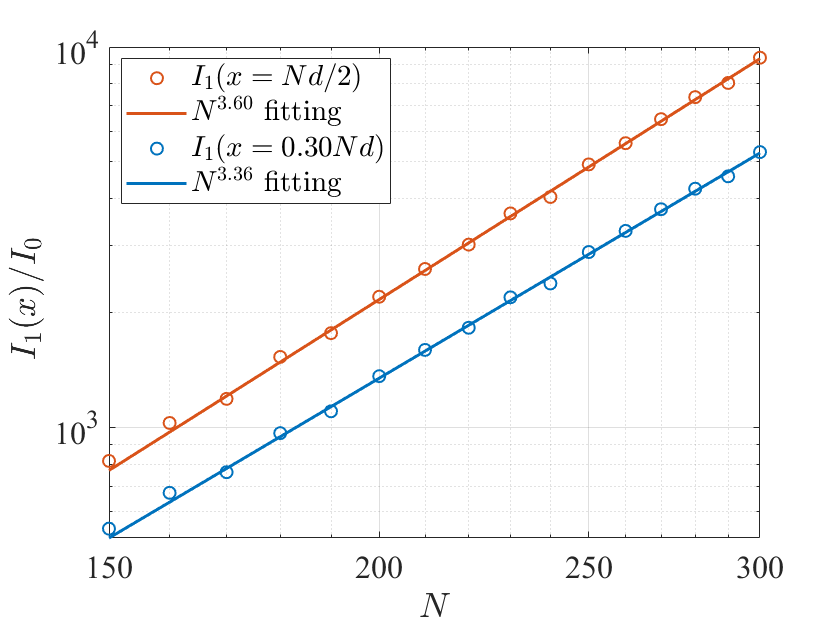}
        \caption{}
        \label{fig:MaxFieldEnhancement_B}
    \end{subfigure}
    \begin{subfigure}[t]{0.45\linewidth}
        \centering
        \includegraphics[width=\linewidth]{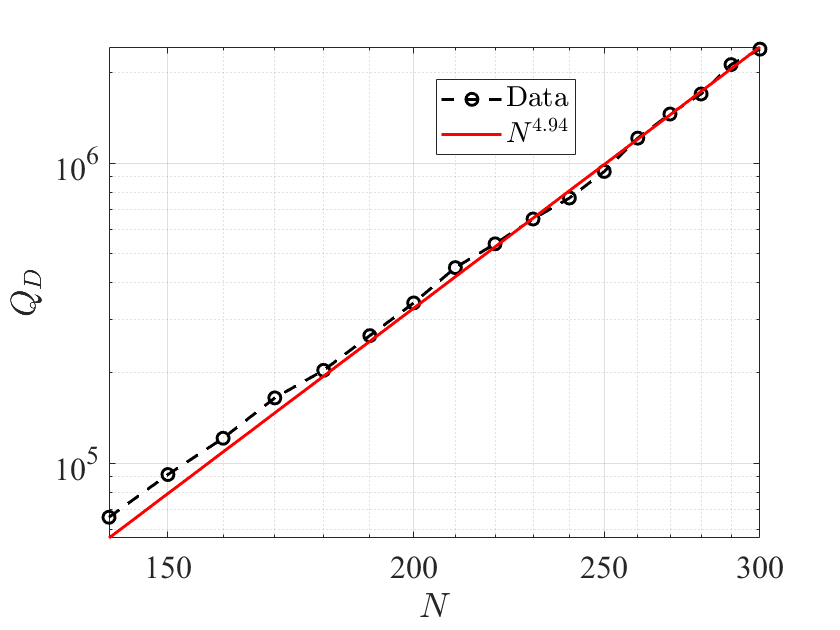}
        \caption{}
        \label{fig:QDvsN}
    \end{subfigure}
    \caption{(a) Difference between the DBE resonance frequency $f_{r,D}$ and the ideal DBE frequency $f_D$ as a function of the number $N$ of unit cells. Black dots are the simulation data; the red line is the fit from Eq.~\eqref{eq:DBEresonFreqAsymp}. (b) Normalized intensity profiles $I_1(x)/I_{1,{\text{max}}}$ of the $z$-polarized field versus longitudinal position $x/(Nd)$ for various waveguide lengths $N$, evaluated at the center of the top waveguide core and at their respective DBE resonances. All normalized profiles collapse onto a single curve, indicating that the spatial distribution of the field is preserved as $N$ increases. This confirms that increasing the number of unit cells enhances the field amplitude without altering the shape of the intensity envelope. The intensity peak approaches $x=Nd/2$ for increasing $N$. (c) Field intensity values $I_1(N)$ normalized by the incident intensity $I_0 = \qty{3.86e8}{\V\squared/\m\squared}$, evaluated at the cavity center $x = Nd/2$ (orange) and at one-third of the length from the center $x = Nd/3$ (blue) as a function of $N$. Solid lines represent power-law fittings of the form $I_{\text{fit}}(N) = aN^b$, yielding exponents of $b = 3.6$ and $b = 3.35$, respectively.  (d) Loaded quality factors of the DBE-supporting double grating (represented by black dots and evaluated at the DBE resonance for each $N$) for varying grating length. The red line is a fitting function that shows the asymptotic $\propto N^5$ scaling of the DBE quality factor $Q_D$ with waveguide length.}
    \label{fig:SchematicDBESHG}
\end{figure*}

The linear properties of the finite-length double grating around the DBE frequency are evaluated using the wave-optics module in COMSOL Multiphysics. The cavity is excited from the left using the even mode of the input waveguide, launched at a distance $L = 1.8\,\upmu$m from the start of the grating. The excitation is a two-dimensional mode with incident power of 1\,$\upmu$W (assuming a waveguide depth of 1\,$\upmu$m). Perfectly matched layers (PMLs) are applied in both the  longitudinal ($x$) and vertical ($y$) directions with sufficient thickness to absorb outgoing waves before they reach the simulation boundaries.

Figure~\ref{fig:ResvsN} confirms the DBE resonance scaling law in Eq.~(\ref{eq:DBEresonFreqAsymp}): the DBE resonances (black dots) approach $f_D = 193.679$\,THz following the $N^{-4}$ scaling trend of Eq.~(\ref{eq:DBEresonFreqAsymp}). For $N > 200$, the cavity exhibits the asymptotic frozen mode behavior. The small value of $h_D$, normalized by $1/d^4$ with $d = 300$\,nm, validates the extreme flatness of the DBE.

Figure~\ref{fig:normIntensityProfile} displays the normalized fundamental-field intensity envelope $I_1$ in the $x$ direction across the cavity of length $Nd$ for different values of $N$. The field intensities are evaluated at the center of the top waveguide core and at their respective DBE resonances. The collapse of all normalized curves into a single profile confirms that the normalized intensity distribution inside the cavity remains nearly unchanged with $N$. 

Importantly for this study, the maximum field intensity enhancement provided by the DBE-associated frozen mode, theoretically should scale asymptotically as $\text{max}(I)\propto N^4$, leading to an extraordinary enhancement of the nonlinear polarization excited at the second-harmonic frequency.  Figure~\ref{fig:MaxFieldEnhancement_B} displays the field intensity at the center of the cavity (as orange dots) and at one third of the waveguide length from the center (as blue dots). The solid lines of the same color are the fitting functions $I_{1,fit}(N)=aN^b$, with $a=4.5\times 10^{12}$, $b=3.6$ at the center of the cavity, and $a=2.5\times 10^{12}$, $b=3.36$ at a third of the cavity length (fit for $N\in[240, 300]$). Both curves are normalized by the incident intensity $I_0 = 3.86\times 10^8 $ $(\text{V/m})^2$. While the exponent of the field enhancement at the  {\em center} of the cavity falls short of the ideal $N^4$ scaling predicted by Eq.(\ref{eq:MaxFieldEnhancement}), it sizably exceeds the $N^2$ intensity growth when operating near an RBE frequency, underscoring the superior field enhancement provided by the DBE. The deviation from $N^4$ scaling is attributed to the mesh discretization and the finite nature of the simulated structure. Field enhancement also varies with the geometry, material platform, and length of the waveguide. Different DBE-supporting designs such as that in Ref.~\cite{nada_giant_2018} exhibit different enhancement factors.

The resulting frozen mode exhibits greater field intensity enhancement and group delay compared to the slow-wave resonance associated with the RBE \cite{figotin_slow_2011}. As a result, waveguides operating near the DBE typically outperform those near an RBE. 

The quality factor for an RBE increases asymptotically with waveguide length as $Q_R \propto N^3$, whereas for a DBE it scales as $Q_D \propto N^5$ \cite{figotin_slow_2011, wood_degenerate_2015, othman_giant_2016, nada_theory_2017, nada_giant_2018}. 

Figure~\ref{fig:QDvsN} shows the quality factor of the cavity at the DBE resonance as a function of the waveguide length (given by $N$). The quality factor is computed for a loaded double grating structure extended continuously on both sides (without mirrors or corrugations) as regular transmission lines with height $h$ as illustrated on the sides of Figures~\ref{fig:SchematicDBESHG_a} and Figure~\ref{fig:double_grating}. The quality factor is extracted using the spectral definition $Q = f_r / \text{BW}$, where BW is the $3$\,dB bandwidth of the transmission peak at resonance frequency $f_r$. This definition aligns well with other methods, as shown in \cite{christopoulos_on_2019}. The quality factors are depicted as black dots, while the solid red line represents the fitting function $Q_{fit} (N)= aN^b$, with $a=2.43\times 10^6$ and $b=4.94$ (fit for $N\in[200,300]$). Within some numerical approximation, this result confirms the expected $N^5$ scaling of the DBE-associated resonance, consistent with prior works \cite{figotin_slow_2011, wood_degenerate_2015, othman_giant_2016, nada_theory_2017, nada_giant_2018, othman_experimental_2017}. The steep $Q \propto N^5$ scaling observed arises from a strong impedance mismatch at the terminations, driven by destructive interference between nearly degenerate propagating and evanescent modes at the DBE resonance \cite{figotin_slow_2011, gutman_degenerate_2011}. This interference necessitates large modal amplitudes to satisfy boundary conditions, resulting in enhanced field amplitudes deep inside the cavity. In contrast, in RBE-based structures the evanescent components are negligible and one has $Q \propto N^3$. Larger $Q$ also imply increased robustness of the DBE mode against termination imperfections and improved SHG performance away from the cavity edges.

In summary, the DBE-supporting double-grating displays a strong scaling of the DBE resonance frequencies $f_{r,D}$, which approach $f_D$ with an $N$-dependent difference that decreases as $N^{-4}$, as predicted by \eqref{eq:DBEresonFreqAsymp}. Additionally, we have shown that the double grating displays a maximum field intensity $I_1$ at the center of the cavity that scales as $N^{3.6}$ and a loaded quality factor that approximately scales as $N^5$ with waveguide length (as opposed to $\text{max}(I)\propto N^2$ and $Q\propto N^3$ when operating near an RBE).

\section{Frozen mode second harmonic generation enhancement}
\label{ch:shg}

The goal of this section is to demonstrate the extraordinary enhancement of the SHG enabled by the DBE and its scaling laws with the grating length. We focus on the second-order nonlinear polarization density component that converts $z$-polarized pump light to $z$-polarized SH light, given by:

\begin{equation}
    P_z(f_2) = \varepsilon_0\frac{2\sqrt{2}}{\sqrt{3}}d_{14} E_z^2(f_1),
    \label{eq:NonlinearPolarization}
\end{equation}

where $d_{14} = 70$ pm/V is the nonlinear coefficient of AlGaAs \cite{chen_proposal_2012} and $E_z$ is the electric field at the fundamental frequency $f_1$. As reported in Appendix A, this nonlinear process can be designed by orienting the crystal with its $[111]$ axis parallel to the $y$-axis and the $[1\bar{1}0]$-axis of the crystal parallel to the $x$-axis. In this way, the polarization, as well as the electric field of the radiated second-harmonic light are also parallel to $z$. This choice ensures that any guided Bloch modes at $f_2$, if present, would be efficiently excited. Below, we show that we have no evidence of guided Bloch mode excitation at $f_2$, confirming the non-resonant character of the SHG emission making the structure less sensitive to material losses at $f_2$. Additional details on the polarization response due to the crystal cut orientation are provided in Appendix~A.

\begin{figure}
    \centering
   \includegraphics[width=0.9\linewidth]{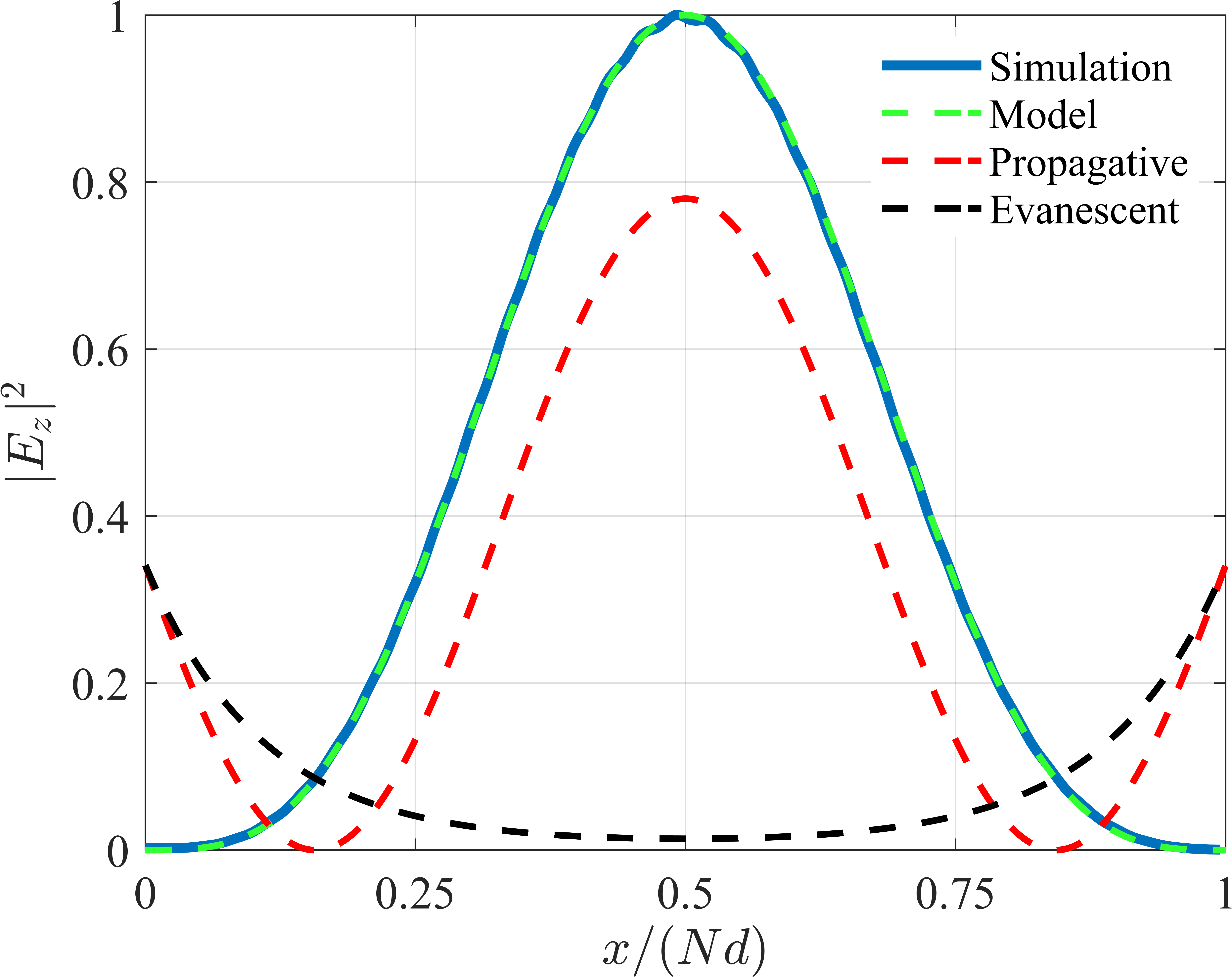}
    \caption{Fundamental-field intensity in the cavity at the DBE resonance, for $N=200$. The Bloch model (green dashed curve, retrieved from Eq. (\ref{pump_field})), shows excellent agreement with the numerical simulation (blue curve). The red-dashed line is the intensity of the propagating portion of the field, $|A_0e^{ik^{(0)}x}+A_2e^{ik^{(2)}x}|^2$, while the black-dashed line is the intensity of the evanescent portion of the field, $|A_1e^{ik^{(1)}x}+A_3e^{ik^{(3)}x}|^2$. All the quantities are normalized with respect to the value of the peak intensity in the center of the cavity.}
    \label{fig:model}
\end{figure}

The energy and momentum conservation conditions \cite{boyd_nonlinear_2008} in a periodic waveguide become

\begin{equation}
    \begin{array}{c}
    f_2 = f_1^+ + f_1^-, \\[1.5ex]
    k_2 = k_1^+ + k_1^-, \;\; 2k_1^+, \;\;  2k_1^-
    \end{array}
    \label{eq:EnergyMomentumConditions}
\end{equation}

where the $\pm$ superscripts refer to the two counter-propagating photons at the fundamental frequency. Since SHG occurs at the resonance of the fundamental frequency, $f_1^+=f_1^-=f_{r,D}$, and their corresponding wavenumbers are $k_1^+ = \pi/d + \chi$ and $k_1^- = \pi/d - \chi$, where $\chi = (\pi-\varphi)/(Nd)$, as derived from Eqs.~(\ref{eq:DBERBEDisp}) and (\ref{eq:DBEresonFreqAsymp}). This implies $k_2 = 2\pi/d$ for all $N$, in addition to $k_2 = 2\pi/d + 2 \chi$  and $k_2 = 2\pi/d - 2 \chi$. For large $N$, these three wavenumbers are all close to $2\pi/d$, and the fundamental spatial harmonic at $k=0$ of such Bloch mode has the proper phase condition for vertical second-harmonic emission. Importantly, the DBE-associated frozen mode at the fundamental frequency $f_{r,D}$ is composed of four waves, rather than the two waves at a resonance near an RBE. In fact, it is the coalescence of those four waves of the frozen mode that provides the enhanced field amplitude away from the cavity edges \cite{figotin_slow_2011, veysi_degenerate_2018}. \\
We now derive approximate analytical expressions of the pump field distribution and the resulting second-harmonic source in terms of Bloch modes. The pump electric field in the finite periodic structure is  expressed as a superposition of the four Bloch modes of the corresponding infinite structure: 
\begin{equation}
E_z(f) = \sum_{0}^{3} A_m e^{ik_1^{(m)} x}, 
    \label{eq:FourModes}
\end{equation}
where the wavenumbers near the DBE take the form $k_1^{(m)}=\pi/d + \chi e^{im \pi/2}$. Here, $k_1^{(0)}=k_1^+$ and $k_1^{(2)}=k_1^-$ are associated with two propagating modes, whereas $k_1^{(1)}$ and $k_1^{(3)}$ are associated with two evanescent modes. Moreover, $\chi=(\pi-\varphi)/(Nd)$ is the distance of the four eigenmodes from the edge of the Brillouin zone, and $\varphi$ accounts for the phase gathered by the propagating modes when reflected at the opposite edges of the cavity \cite{burr_degenerate_2013}. The amplitudes $A_m=A_m(x,y)$ are periodic in $x$ with period $d$. To obtain a simplified representation near the DBE, we approximate the $A_m$ as constants, focusing on one of the Floquet harmonics. At the DBE resonance of the finite structure, the total field is much smaller at the edges of the cavity than in its center, and we know that this phenomenon is due to the strong excitation of evanescent waves at either end of the cavity \cite{figotin_gigantic_2005}. In other words, the DBE resonance requires that the intensity of the propagating portion of the field be approximately equal to the intensity of the evanescent portion of the field, with the two almost canceling each other at the two interfaces $x=0$ and $x=Nd$. In addition, the two counter-propagating components should have approximately equal amplitudes and interfere constructively at the center of the cavity. With these prescriptions in mind, we obtain $A_1\approx-A_0 [1+e^{i(\pi-\varphi)}]/[1+e^{-(\pi-\varphi)}] $, $A_2\approx A_0e^{i(\pi-\varphi)}$, and $A_3\approx A_1e^{-{(\pi-\varphi)}}$. It follows that the pump electric field simplifies to an expression that depends on $\varphi$ and $A_0$,

\begin{align}
\label{pump_field}
E_z(f_1) &= 2 A_0 e^{i\pi x/d} e^{i(\pi-\varphi)/2} \biggl[  \sin\left(\chi x+\varphi/2\right) +\notag \\
       &\quad -\frac{\sin \left( \varphi/2 \right)}{\cosh \left(\frac{\pi-\varphi}{2} \right)} \cosh\left(\chi x-\frac{\pi-\varphi}{2}\right) \biggl],
\end{align}

highlighting the presence of a standing wave formed by the counter-propagating Bloch modes, with a maximum intensity at $x=Nd/2$, as well as the presence of an amplitude term originated by the two evanescent Bloch modes, i.e., the $\cosh$ term. We stress that a DBE resonance is only possible if $\varphi\neq0$. In fact, for $\varphi=0$, the amplitudes of the evanescent portion of the field would be equal to zero and the field in Eq.~\ref{pump_field} would reduce to the typical shape of an RBE cavity. Instead, setting the single fitting parameter to $\varphi=-0.46 \pi$ yields good agreement of the Bloch model with the numerical simulation, as illustrated in Fig.~\ref{fig:model}. The Bloch decomposition allows the isolation of the contribution of the propagative and the evanescent components of the field. The intensity of these components is reported in Fig.~\ref{fig:model}, along with the total field intensity. The evanescent and propagative portions of the field cancel out at the edges of the cavity, while at the center of the cavity all four Bloch modes add in phase. It is important to notice that the evanescent modes add a non-negligible contribution to the overall pump intensity peak at the cavity center. Indeed, the pump intensity at $x=Nd/2$ can be written as

\begin{equation}
    \label{peak_intensity}
    I_{1,\text{max}}=4|A_0|^2 \left[ 1-\frac{\sin (\varphi/2)}{\cosh(\frac{\pi -\varphi}{2})} \right]^2,
\end{equation}

With $\varphi=-0.46 \pi$, the parenthesis in Eq. (\ref{peak_intensity}) evaluates to $\approx 1.28$, which means that the evanescent portion of the field is providing almost a $30\%$ increase of pump intensity, which further boosts the nonlinear processes beyond having a large $A_0$. This enhancement is a hallmark of a DBE cavity. In an RBE cavity, where $\varphi=0$, only the two counterpropagating Bloch modes interfere constructively at the cavity center, producing a $\sin^{2}$ intensity profile and a smaller coefficient $A_0$.

The $z$-component of the nonlinear polarization density induced in the DBE cavity at the second-harmonic frequency, 

\begin{align}
&P_z(2f_1) \propto 4 A_0^2 \,e^{i\frac{2\pi}{d}x} e^{i(\pi-\varphi)}\Biggl[ \sin^2 \left( \chi x+\varphi/2\right)+ \notag \\
&\quad +\frac{\sin^2 \left( \varphi/2\right)}{\cosh^2 \left( \frac{\pi -\varphi}{2} \right)} \cosh ^2 \left( \chi x -\frac{\pi - \varphi}{2}\right)+\notag \\
&\quad -\frac{2\sin \left( \varphi/2\right)}{\cosh \left( \frac{\pi -\varphi}{2} \right)} \cosh  \left( \chi x -\frac{\pi - \varphi}{2}\right)\sin \left( \chi x +\phi/2\right)\Biggr], 
\label{SH_source}
\end{align}

reveals: (i) the ability of the structure to emit second-harmonic light in the vertical  direction (i.e., normal to the longitudinal grating length), due to the fact that the SHG polarization density source $P_z$ shows the same phase in each period of the structure; (ii) the crucial role of the evanescent modes supported by the DBE cavity in shaping and increasing the overall intensity of the nonlinear polarization density source.

As shown in Figure~\ref{fig:Dispersions}, the DBE has a bandgap below $f_D$, so the fundamental frequency satisfies $f_1 \equiv f_{r,D}> f_D$, with $f_1 \to f_D$ only for $N \to \infty$. The dispersion at the second harmonic, shown in the inset in Figure~\ref{fig:Dispersions}, confirms that no Bloch modes exist near $\operatorname{Re}(k d) = 2\pi$ at $f_2 \approx 2f_D$. Thus, the conditions in Eq.~\eqref{eq:EnergyMomentumConditions} are not met, and the nonlinear polarization does not couple to propagating modes at $f_2$. The observed SHG is therefore entirely attributed to the non-phase-matched nonlinear effect induced by the frozen mode at $f_1$.

Operating the pump at the DBE resonance $f_1 =f_{r,D}$ implies a second-harmonic frequency of $f_2 = 2f_{r,D}$, which asymptotically follows

\begin{equation}
f_{2} \sim 2f_{D} + \frac{h_D}{ \pi} \left(\frac{\pi-\varphi}{N d}\right)^4,
\label{eq:SeconHFreqAsymp}
\end{equation}

as $N$ increases. Since no guided Bloch modes exist at $f_2=2f_{r,D}$, the SHG radiates vertically through the top and bottom cladding. Due to the strong impedance mismatch at the cavity terminations at the DBE resonances (discussed in Section~\ref{ch:FF}), the second-harmonic polarization density source is concentrated near the center of the cavity and suppressed at its terminations, following the $I_1^2$ pattern in Figure~\ref{fig:normIntensityProfile}.

\begin{figure*}[ht]
    \centering
    \begin{subfigure}[t]{0.45\linewidth}
        \centering
        \includegraphics[width=\linewidth]{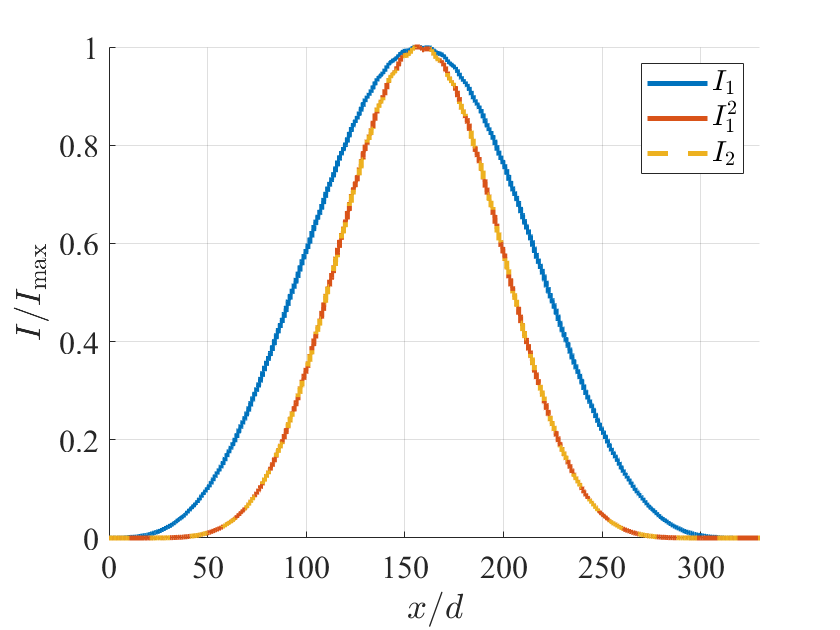}
        \caption{}
        \label{fig:intensityProfiles}
    \end{subfigure}
    \begin{subfigure}[t]{0.45\linewidth}
        \centering
        \includegraphics[width=\linewidth]{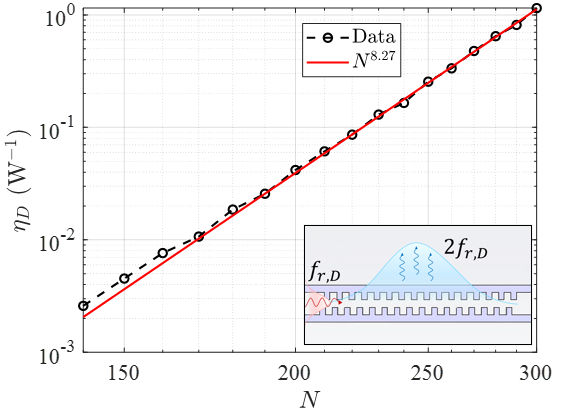}
        \caption{}
        \label{fig:EtavsN}
    \end{subfigure}
    \caption{(a) Normalized intensity profiles along the longitudinal $x$ axis of a double grating with $N=330$ unit cells at the center of the top waveguide core. The solid blue line shows the intensity $I_1$ at the fundamental frequency $f_1 = f_{r,D}$, while the solid red line represents its squared value, $I_1^2$. The dashed yellow line denotes the normalized intensity $I_2$ at the second harmonic ($f_2 = 2f_{r,D}$). The near-perfect overlap between $I_1^2$ and $I_2$ confirms that the second-harmonic field distribution follows the nonlinear polarization generated by the DBE-enhanced field, and that no guided Bloch modes are substantially excited at $f_2$. (b) Conversion efficiency of double grating (black dots) cavities of different lengths computed at their respective DBE resonances. The solid red line depicts the fitting function that shows the asymptotic $N^{8.27}$ scaling of the nonlinear conversion efficiency $\eta_D$ with waveguide length. The inset depicts the field at the fundamental frequency $f_1=f_{r,D}$ being excited from the left and the field at the second-harmonic frequency leaking out of the structure.}
    \label{fig:CombinedSHGResults}
\end{figure*}

Figure~\ref{fig:intensityProfiles} depicts the normalized field intensity $I = |\mathbf{E}|^2$ along the cavity (evaluated at the middle of the top waveguide core) for a structure with $N = 330$. The intensity profile $I_1$ at the fundamental frequency $f_1=f_{r,D}$ (the DBE resonance) is shown as a solid blue curve, while its squared value $I_1^2$ is plotted as a solid red curve. The dashed yellow curve represents the normalized intensity $I_2$ at the second-harmonic frequency $f_2=2f_{r,D}$. The near-perfect agreement between $I_{1}^2$ and $I_{2}$ in Figure~\ref{fig:intensityProfiles} and the results shown in Figure~\ref{fig:Dispersions} are evidence that the second-harmonic field arises mostly from the nonlinear polarization density driven by the DBE-enhanced fundamental field. Although periodic structures can, in general, enable the excitation of Bloch modes via field gradient, our results show no significant evidence of such excitation.

We define the nonlinear conversion efficiency $\eta$ \cite{duchesne_second_2011, may_second_2019} as

\begin{equation}
    \eta = \frac{P_{2}}{P_1^2},
    \label{eq:DefConversionEfficiency}
\end{equation}

where $P_1$ is the incident power at $f_1=f_{r,D}$ and $P_{2}$ is the output power at $f_2=2f_{r,D}$.  The $P_1^2$ normalization ensures that $\eta$ is independent of the incident power in the undepleted pump approximation \cite{mobini_algaas_2022}. Since we perform 2D simulations for a structure invariant along $z$, here we assume a nominal width of $w=1$ $\upmu$m, as shown in Figure~\ref{fig:SchematicDBESHG_a} and the incident power $P_1$ over that width is expressed in units of W. As discussed next, $P_2$ is the power extracted vertically, and it has also a unit of W; hence the efficiency has units of W$^{-1}$.

The SHG power radiated vertically is calculated as

\begin{equation}
    P_{2} = \int_{0}^{Nd}  S_y(x)  \ dx,
    \label{eq:PointingIntegral}
\end{equation}

where $S_y(x)$ denotes the vertical component of the time-averaged Poynting vector $\mathbf{S}=\operatorname{Re}(\mathbf{E}\wedge\mathbf{H}^*)/2$. This integration is performed at a distance equal to one period $d$ from the end of the top waveguide core to allow the field of higher-order Floquet spatial harmonics to decay.

The nonlinear properties of the finite-length double gratings at the DBE resonances are obtained using the frequency-domain solver of the wave-optics module in COMSOL Multiphysics. The structure is excited at the $N$-dependent DBE resonance frequency $f_1  \equiv  f_{r,D}$ in the positive $x$ direction, as in the linear studies in Section~\ref{ch:FF}, inducing the nonlinear polarization described by Eq.~\eqref{eq:NonlinearPolarization}. The same boundary conditions used in the linear simulations are applied.

The main result of this paper is the asymptotic scaling of the nonlinear conversion efficiency $\eta$ with the waveguide length, enabled by the exceptional light--matter interaction between the DBE-associated frozen mode and the nonlinear medium in the waveguide core. Figure~\ref{fig:EtavsN} displays $\eta$ as a function of the number of unit cells $N$, depicted as black dots. The inset shows the field at the fundamental frequency excited from the left whereas that at the second harmonic at $f_2\equiv 2 f_{r,D}$ is extracted vertically from the structure. The solid red line represents the fitting function $\eta = aN^b$. The parameters have been evaluated as $a=1.12$ and $b=8.27$, with a fit in the interval $N\in[200,300]$. This exceptional scaling of the nonlinear conversion efficiency is striking compared to what has been published in the literature so far. It is higher than the scaling $\eta \propto N^8$ reported in \cite{deAngelis_Maximum_2007}, found in a doubly resonant structure. The technique reported in \cite{deAngelis_Maximum_2007} was based on exciting an RBE resonance at $f_1$ and having $f_2$ coincident with an RBE resonance at $f_2$, and despite representing a record high scaling of the efficiency with the number of unit cells of the photonic crystal there employed, it requires a precise condition of double resonance at the two RBEs that is not easy to realize. Indeed, the numerical study in \cite{deAngelis_Maximum_2007} was based on a multilayered photonic crystal where the double resonance condition was found assuming zero dispersion of the materials' parameters.

In our case instead, the scaling is obtained just by choosing $f_1$ corresponding to the DBE resonance frequency, without requiring the excitation of any Bloch modes at the second harmonic. The DBE-enhanced steep power scaling arises from the proportionality of the nonlinear conversion efficiency to the square of the field intensity $I_1$, whose maximum has been shown to scale with the waveguide length as $\text{max}(I_1)\propto N^{3.6}$ at the DBE resonances. Since the integral domain of the Poynting vector in Eq.~(\ref{eq:PointingIntegral}) approximately adds an additional $N$ contribution to the vertically leaked SHG power,
\begin{equation}
P_2(N) \propto N  \max(I_2) \propto N  \max(I_1^2)   
\end{equation}
leading to the observed asymptotic law  
$P_2(N) \propto N^{8.2}$.  As a consequence, our numerical findings return a leaked SHG efficiency that scales as $\eta_D\propto N^{8.27}$. These minor discrepancies can be attributed to numerical discretization effects and the difference in the evaluation region: $P_2$ is computed one period above the waveguide well into the cladding, whereas the peak field intensity is taken at the center of the top waveguide core.

The main implications of this trend are that (i) relying on the DBE for nonlinear interactions (rather than on an RBE) allows for a remarkable miniaturization of photonic integrated nonlinear sources and devices beyond previous estimates, even without requiring collinear phase-matching; (ii) despite the DBE being a condition that requires precise fabrication, our proposed technique only requires the tuning of $f_1$ with the resonance frequency of the DBE cavity and no collinear phase matching is required at $f_2$; (iii) in the present double-grating configuration, the SHG field radiates vertically through the cladding, while the fundamental mode is side-coupled, which may simplify device integration. Note that while the second harmonic has been shown not to couple to guided Bloch modes, edge-induced effects could still excite weakly guided or radiative components at $f_2$, although this is not observed due to the high degree of matching between the shape of $I_2$ and that of $I_1^2$.

\begin{widetext}
\begin{table}[ht]
\centering
\resizebox{\linewidth}{!}{%
\begin{tabular}{|l|l|l|c|r|r|c|c|}
\hline
\textbf{Ref.} & \textbf{Material} & \textbf{Device} & Length [mm] & $P_1$ [W] & $P_{2}$ [W] & $\eta = P_2/(P_1^2)$ [W$^{-1}$] & $\eta_{\mathrm{norm}} = P_2/(P_1^2L^2)$ [W$^{-1}$cm$^{-2}$] \\
\hline
\cite{yu_efficient_2005} & AlGaAs & WG & 5 & $1.9\times10^{-3}$ & $0.85\times10^{-6}$ & $0.2$ & 0.83 \\
\cite{roland_second_2020} & AlGaAs & Suspended wire & 1000 & $8\times10^{-4}$ & $3.9\times10^{-9}$ & $0.006$ & $6.1\times 10^{-7}$ \\
\cite{roland_second_2020} & AlGaAs & Suspended rib & 200 & $4.0\times10^{-4}$ & $2.7\times10^{-11}$ & $1.68\times 10^{-4}$ & $4.22\times 10^{-7}$\\
\cite{stanton_efficient_2020} & GaAs & Waveguide & 2.9 & $5\times10^{-6}$ & $1\times10^{-9}$ & 40 & 475.62 \\
\cite{xu_device_1997} & AlGaAs & Waveguide & 0.5 & $68\times10^{-6}$ & $0.12\times10^{-9}$ & 0.026 & 10.38 \\
\cite{abolghasem_highly_2009} & AlGaAs & Waveguide & 2.17 & $3.3\times10^{-3}$ & $60\times10^{-6}$ & 5.5 & 117 \\
\cite{duchesne_second_2011} & AlGaAs & Nano wire & 12.6 & $155\times10^{-6}$ & $50\times10^{-12}$ & 0.002 & 0.001 \\
\cite{savanier_large_2011} & AlGaAs & Waveguide & 0.5 & $0.3\times10^{-3}$ & $2.25\times10^{-6}$ & 25 & $10\times 10^3$ \\
This work & AlGaAs & DBE Double grating & 0.09 & $1 \times 10^{-6}$ & $1.15\times 10^{-12}$ & $1.15$ & $14\times 10^3$ \\
\hline
\end{tabular}%
}
\caption{Comparison of SHG efficiency in various experimental AlGaAs and GaAs waveguides. \\ * To compare the 2D double grating efficiencies, we have assumed a width $w=1$ $\upmu$m.
($N$=300)*}
\label{tab:shg-efficiency}
\end{table}
\end{widetext}

For the DBE-supporting double grating with $N = 300$ unit cells (corresponding to a total length of $L = 0.9$ $\upmu$m), we obtain a second-harmonic generation efficiency of $\eta = 1.15 $ (W)$^{-1}$. Put differently, the structure converts a $4$\% of the pump power into vertically emitted second-harmonic light while operating at an input, peak-power density of $10$~\text{MW/cm}$^2$, a relatively low intensity level in the realm of nonlinear optics. Note that we have assumed a width $w=1$ $\upmu$m. A comparison with previously reported experimental AlGaAs devices is provided in Table~\ref{tab:shg-efficiency}. While the resulting efficiency is moderate compared to the highest values reported in the 7\textsuperscript{th} column of Table~\ref{tab:shg-efficiency}, this comparison does not consider differences in device length. Here we normalize the SHG efficiency by $L^2$ to remove the dependence on the interaction length, as in \cite{abolghasem_highly_2009, duchesne_second_2011, roland_second_2020}. Then, our structure yields $\eta_{\text{norm}} = 14\times 10^3$ W$^{-1}$cm$^{-2}$, which surpasses all the values in Table~\ref{tab:shg-efficiency}. This result highlights the potential for extraordinary miniaturization of nonlinear devices operating near a DBE. Additionally, these results are obtained with no need to optimize the design for overlap between the fundamental and second harmonic modes, since no Bloch modes are excited at the second-harmonic frequency.

The demonstrated grating design also delivers a blueprint for an efficient integrated source of entangled photons. Frequency-degenerate spontaneous parametric down-conversion \cite{kwiat_new_1995}, which is the inverse process of SHG under time reversal, allows free-space pumping at $\lambda_{\mathrm{pump}}=775$\,nm to resonantly generate waveguide-coupled pairs of telecom photons, opening the use of DBEs in quantum optics and secure communications \cite{bogaerts_sillicon_2012}.

\section{Conclusion}
\label{ch:Conclusions}

We have conceived and analyzed a nonlinear grating that supports a degenerate band edge (DBE), a fourth-order exceptional point of degeneracy, as a means to enhance vertical emission of second-harmonic generation (SHG) without exciting any Bloch modes at the second-harmonic frequency. Operating the mirrorless double grating close to the DBE frequency brings about very large field intensities $I_1$ at the DBE resonance, whose maximum scales as $N^{3.6}$, close to the $N^4$ dependence expected from the literature. Additionally, we have shown the asymptotic $N^{4.94}$ quality factor scaling with waveguide length for a large number of unit cells $N$ up to 300. This behavior is a direct result of the DBE-associated frozen-mode regime.

Building on this linear response, we have demonstrated that the resulting field enhancement leads to efficient SHG. The second-harmonic field, which leaks vertically out of the structure, has been shown to follow the same spatial distribution as the square of the fundamental field. Our simulations show that the DBE-enhanced nonlinear conversion efficiency asymptotically scales as $\eta \propto N^{8.27}$ for large $N$, highlighting the effectiveness of the DBE mechanism in boosting light--matter interactions, even without collinear phase matching and in the absence of Bloch modes at the second-harmonic frequency. Additionally, when the efficiency is normalized by the square of the waveguide length, the DBE-supporting double grating efficiency is predicted to surpass the state-of-the-art, demonstrating a strong potential for the miniaturization of integrated photonic nonlinear sources.

These results establish the DBE as a powerful mechanism for compact and efficient nonlinear integrated photonic devices. The structure proposed here is compatible with existing material platforms and provides a robust, scalable approach for boosting nonlinear effects in applications ranging from on-chip frequency conversion to quantum photonic systems.

\section*{Acknowledgment}
A.Z. acknowledges the U.S.–Italy Fulbright Commission for supporting his stay at the University of California, Irvine through a Visiting Research Scholar grant.

\section*{Disclosure}
The authors declare no conflicts of interest.

\section*{Data availability statement}
Data underlying the results presented in this paper are not publicly available at this time but may be obtained from the authors upon reasonable request.

\section*{Appendix A: Nonlinear polarization sources}
\label{ch:Appendix}

In the AlGaAs-based waveguide, the relation between the polarization density induced at the second harmonic, $\mathbf{P}(2f)$, and the electric field at the fundamental frequency, $\mathbf{E}(f)$, depends on the orientation of the exciting polarization with respect to the crystal lattice.

For a zincblende crystal (point group $\bar{4}$3m), such as AlGaAs in our waveguide, the elements of second-order nonlinear susceptibility tensor are null if any two indices are equal, namely, in terms of Kronecker delta
\begin{equation}
    \label{eq:zincblende_chi2}
    \chi^{(2)\prime}_{ijk}
=   2d_{14}(1-\delta_{ij})(1-\delta_{ik})(1-\delta_{jk}).
\end{equation}

%
%

This known expression holds in the reference frame of the crystal axes ($\hat{x}^\prime,\hat{y}^\prime,\hat{z}^\prime$).
In our waveguide design, the crystal axis [111] is parallel to the $y$-axis of the laboratory frame ($\hat{x},\hat{y},\hat{z}$), i.e, the vertical direction indicated in Figure~\ref{fig:SchematicDBESHG_a}, while $\hat{x} \equiv [1\bar{1}0]$ and $\hat{z}=\hat{x}\wedge\hat{y} \equiv  [\bar{1}\bar{1}2]$.
The base unit vectors of the laboratory frame form the columns of the rotation matrix $\bm{R}$ converting from the crystal to laboratory frame (namely $\bm{R}\mathbf{v}^\prime = \mathbf{v}$ for a generic vector $\mathbf{v}$)
\begin{equation}
    \label{eq:rotation_matrix}
\bm{R}    =
\begin{bmatrix}
\frac{1}{\sqrt{2}} & \frac{1}{\sqrt{3}} & -\frac{1}{\sqrt{6}}\\
-\frac{1}{\sqrt{2}} & \frac{1}{\sqrt{3}} & -\frac{1}{\sqrt{6}}\\
0 & \frac{1}{\sqrt{3}} & \frac{2}{\sqrt{6}}
\end{bmatrix}
    =\frac{1}{\sqrt{6}}
\begin{bmatrix}
 \sqrt{3} & \sqrt{2} & -1\\
-\sqrt{3} & \sqrt{2} & -1\\
0         & \sqrt{2} &  2\\
\end{bmatrix}.
\end{equation}
The nonlinear susceptibility tensor transforms according to
\begin{equation}
    \chi^{(2)}_{lmn}
=   \sum_{i,j,k}R_{il}R_{jm}R_{kn}\chi^{(2)\prime}_{ijk}
\end{equation}
where one has to plug in Eq.~\eqref{eq:zincblende_chi2} and Eq.~\eqref{eq:rotation_matrix}.
For SHG, $\chi^{(2)}$ can always be recast in the compact notation, that in the laboratory frame reads 
\begin{equation}
    \bm{d} = \frac{d_{14}}{3}
    \begin{bmatrix}
        0 & 0 & 0 & 0 & -\sqrt{6} & -\sqrt{3}  \\
        -\sqrt{3} & 2\sqrt{3} & -\sqrt{3} & 0 & 0 & 0  \\
        -\sqrt{6} & 0 & \sqrt{6} & -\sqrt{3} & 0 & 0 
    \end{bmatrix}.
\end{equation}
In terms of $\bm{d}$, the nonlinear polarization sources are
\begin{equation}
    \begin{bmatrix}
        P_x(2f) \\
        P_y(2f) \\
        P_z(2f)
    \end{bmatrix} = 2 \varepsilon_0 \bm{d} \begin{bmatrix}
        E_x^2(f) \\
        E_y^2(f) \\
        E_z^2(f) \\
        2E_y(f)E_z(f) \\
        2E_x(f)E_z(f) \\
        2E_x(f)E_y(f)        
    \end{bmatrix}.
\end{equation}
In our waveguide, the pump guided mode is TE-polarized, i.e., the only non-zero component of the pump electric field is $E_z(f)$. Therefore, the nonlinear sources reduce to
\begin{equation}
    \begin{bmatrix}
        P_x(2f) \\
        P_y(2f) \\
        P_z(2f)
    \end{bmatrix} =2\varepsilon_0 \frac{d_{14}}{\sqrt{3}} \begin{bmatrix}
        0 \\
        -1 \\
        \sqrt{2}
    \end{bmatrix}E_z^2(f)
\end{equation}
which shows that the chosen crystal orientation allows to generate a $z$-polarized SH field with a $z$-polarized pump via the source component $P_z(2f)$.





\section*{Appendix B: Numerical Methods}
\label{sec:NumericalMeth}

The dispersion curves of the periodic structure are obtained by computing the eigenmodes using the eigenmode solver in COMSOL Multiphysics applied to a single unit cell.

To reduce the computational cost, we perform 3D simulations with appropriate boundary conditions. Perfect-electric-conductor (PEC) boundaries are imposed on two nearby planes normal to the $z$-direction, which preserves the essential modal physics of the coupled-grating system while limiting the computational domain. Perfectly matched layers (PMLs) are applied in the vertical ($y$) direction, and the cladding thickness is chosen such that the fields decay sufficiently before reaching the PML boundaries. Floquet periodic boundary conditions are imposed on the two boundaries of the unit cell (i.e., the two faces normal to the $x$ direction). This configuration suppresses spurious reflections from artificial boundaries and ensures proper absorption of outgoing radiation, thereby mimicking an open and infinite simulation volume.

\begin{figure}
    \centering
    \includegraphics[width=0.8\linewidth]{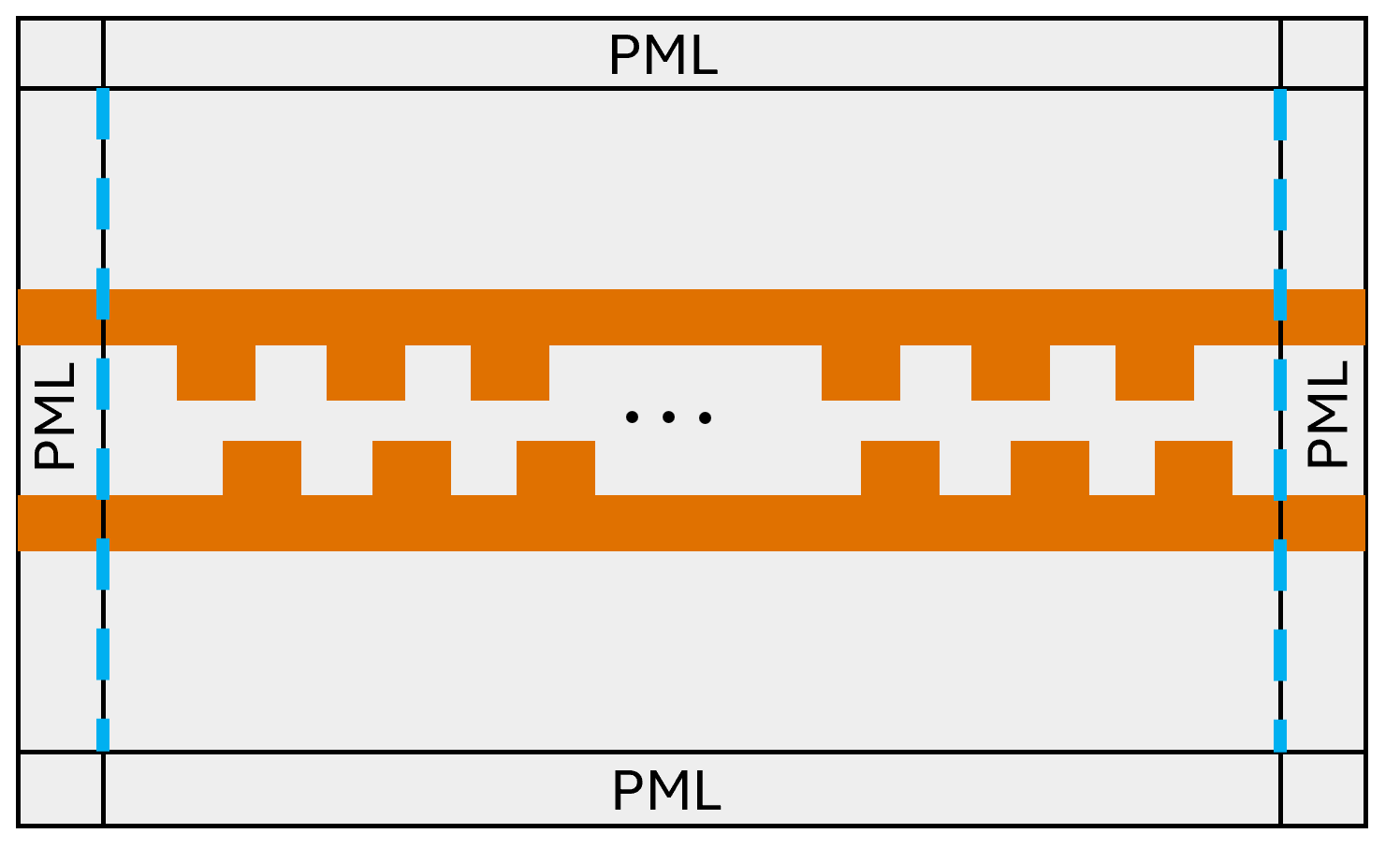}
    \caption{Setup of the numerical experiment for second-harmonic generation. At the fundamental frequency, input and output waveguide ports (blue dashed lines) are placed in the unperturbed waveguide sections. PMLs surround the structure to absorb scattered light. The second-harmonic simulation uses the same geometry and absorbing boundaries, but ports are removed and nonlinear polarization density sources are introduced in the high-index nonlinear regions.}
    \label{fig:whole_structure}
\end{figure}

Second-harmonic generation (SHG) is computed by first solving the electromagnetic problem at the fundamental frequency. The simulated structure consists of three sections, as shown in Fig.~\ref{fig:whole_structure}: (i) an input section on the left, composed of two coupled and unperturbed waveguides; (ii) a cavity section consisting of $N$ periods of the double grating; and (iii) an output section identical to the input section. The structure is excited from the left by launching the first even mode of the two straight input waveguides, at a distance $L = 1.8\,\upmu$m from the beginning of the grating. The excitation is implemented as a port mode (a 2D transverse field) with an incident power of $1\,\upmu$W (assuming a waveguide depth along $z$ of $1\,\upmu$m). The input port is backed by a PML to suppress reflections. A similar configuration is used in the output section, where a receiving port (with zero injected power) is placed and backed by an absorbing PML. PMLs are also applied in the vertical ($y$) direction, as shown in Fig.~\ref{fig:whole_structure}. This configuration enables absorption not only of the power carried by the dominant guided mode, but also of radiation leaking into the cladding and power coupled into higher-order modes of the external waveguides.

The electromagnetic problem at the second-harmonic frequency is solved using the same PML configuration in the $x$ and $y$ directions, but without ports. The source term in the second-harmonic simulation is the nonlinear polarization density induced in the AlGaAs regions by the fundamental-frequency field.

\label{ch:Appendix}

\bibliographystyle{ieeetr}
\bibliography{MAIN_V1}

\end{document}